# Nonlinear optical imaging of in-plane anisotropy in two-dimensional SnS

G. M. Maragkakis[1,2+], S. Psilodimitrakopoulos[1+], L. Mouchliadis[1] A. S. Sarkar[1], A. Lemonis[1], G. Kioseoglou[1,3], and E. Stratakis[1,2*]

[1]Institute of Electronic Structure and Laser, Foundation for Research and Technology-Hellas, Heraklion Crete 71110, Greece

[2]Department of Physics, University of Crete, Heraklion Crete 71003, Greece

[3]Department of Materials Science and Technology, University of Crete, Heraklion Crete 71003, Greece

[*] Correspondence: E. Stratakis, e-mail: **stratak@iesl.forth.gr**

[+] These authors contributed equally to this work

**Abstract**

Two-dimensional (2D) tin(II) sulfide (SnS) crystals belong to a class of orthorhombic semiconducting materials that are lately attracting significant interest, given their remarkable properties, such as in-plane anisotropic optical and electronic response, multiferroic nature and lack of inversion symmetry. The 2D SnS crystals exhibit anisotropic response along the in-plane armchair (AC) and zigzag (ZZ) crystallographic directions, offering an additional degree of freedom in manipulating their behavior. Therefore, calculating the AC/ZZ directions is important in characterizing the 2D SnS crystals. In this work, we take advantage of the lack of inversion symmetry of the 2D SnS crystal, that produces second harmonic generation (SHG), to perform polarization-resolved SHG (P-SHG) nonlinear imaging of the in-plane anisotropy. We fit the P-SHG experimental data with a nonlinear optics model, that allows us to calculate the AC/ZZ orientation from every point of the 2D crystal and to map with high-resolution the AC/ZZ direction of several 2D SnS flakes belonging in the same field of view. It is found that the P-SHG intensity polar patterns are associated with the crystallographic axes of the flakes and with the relative strength of the second order nonlinear susceptibility tensor in different directions. Therefore, our method provides quantitative information of the optical in-plane anisotropy of orthorhombic 2D crystals, offering great promise for performance characterization during device operation in the emerging optoelectronic applications of such crystals.

# 1. Introduction

Group IV monochalcogenides, also known as group IV-VI metal monochalcogenides, and denoted by MX with M = Sn, Ge and X = S, Se, are a class of layered, orthorhombic, semiconducting 2D materials attracting significant interest [1-3]. They are known as phosphorene analogues [1-4], since they share similar puckered or wavy lattice structures with phosphorene, a 2D format of black phosphorus [5, 6]. The in-plane structural anisotropy of MXs, with puckered structure along the AC direction [3], is the origin of in-plane anisotropic physical properties [1-3, 7, 8]. A plethora of properties have been reported to exhibit in-plane anisotropic response, including carrier mobility [7], optical absorption, reflection, extinction, refraction [8], and Raman spectral behavior [1]. The in-plane anisotropic response is exhibited along the distinguished in-plane AC and ZZ crystallographic directions, offering an additional degree of freedom in manipulating their properties [1-3, 7, 8]. For example, polarization-sensitive photodetectors based on the intrinsic linear dichroism of GeSe [8] and black phosphorus [9] have been presented. Furthermore, monolayers MXs are predicted to be multiferroic with coupled ferroelectricity and ferroelasticity, and large spontaneous polarization [10, 11]. Indeed, in-plane ferroelectricity has been recently demonstrated for micrometer-size monolayer SnS at room temperature [12].

Recently, the nonlinear optical properties of MXs have been addressed [12-16]. Nonlinear optics plays an important role in all aspects of modern photonics, with nonlinear media being used in photonic devices for photon generation, manipulation, transmission, detection and imaging [17-20]. Applications of nonlinear optics in a wide range of fields have been explored, including nonlinear silicon photonics [21], quantum nonlinear optics [22], nonlinear plasmonics [23], material characterization [24-26] and biomedical optics [27, 28]. SHG is possibly the most widely studied nonlinear optical process, in which radiation at twice the frequency of the incident light is generated [17-20]. It emerges in media that lack inversion symmetry, such as various 2D layered materials, and is widely used to characterize their properties [24-26, 29, 30]. Furthermore, SHG has been combined with microscopy techniques enabling imaging of 2D materials. In this context, P-SHG imaging has been recently demonstrated as a powerful tool to probe the properties of 2D group VI transition metal dichalcogenides (TMDs), such as $MoS_2$, $WS_2$, $MoSe_2$ and $WSe_2$ [31-36]. Specifically, it has been used to calculate in a pixel-by-pixel manner the main crystallographic direction (armchair) of 2D TMDs and quantify their crystal quality [31, 32], determine the twist-angle in TMD homobilayers [33, 34] and heterobilayers [35], and probe the valley population imbalance [36].

In this work, we extend the use of the P-SHG microscopy technique in order to investigate the properties of the orthorhombic 2D MXs. The 2D MXs are characterized by broken inversion symmetry, a fact that renders them suitable for SHG conversion [12-15]. Indeed, using first-principles electronic structure theory, Wang and Qian theoretically predicted giant optical SHG in monolayer MXs [13]. They predicted that the strength of SHG susceptibility of GeSe and SnSe monolayers is more than one order of magnitude higher than that of monolayer $MoS_2$. These results were also supported by another theoretical work by Panday and Fregoso [14]. Recently, Higashitarumizu et al. performed polarized SHG spectroscopy on a micrometer-size monolayer SnS [12], while Zhu et al. reported anisotropic SHG in few-layer SnS [15].

The P-SHG methodology applied here is based on high-resolution P-SHG imaging microscopy, with spatial resolution of approximately 400nm (see Methods). The subsequent fitting of the P-SHG polar diagrams for every pixel of the image with a theoretical model that accounts for the orthorhombic crystal structure of MXs, enables the calculation of the AC/ZZ direction from every point of a 2D SnS flake and the estimation of two ratios of the second-order nonlinear optical susceptibility tensor elements. We perform the same procedure for several different 2D SnS crystal flakes within the same field of view. It is shown that the mean and the standard deviation of the spatial distributions of the acquired values provide new means of contrast capable to discriminate 2D SnS crystals in the same image based on their in-plane structural anisotropy. Therefore, our technique provides insight into the nonlinear optical properties of 2D MXs and can serve as a useful characterization tool for emerging applications.

## 2. Methods

### 2.1 SHG from orthorhombic MXs

In order to describe the interaction of an excitation field with a 2D orthorhombic MX crystal (Fig. 1a) and the subsequent production of SHG, we use the Jones formalism [31-37]. The two coordinate systems considered are schematically shown in Fig. 1b: the laboratory frame (X, Y, Z) and the crystal coordinates (x, y, z), where $z \equiv Z$. The laser beam propagates along Z axis, normally incident on the crystal, and linearly polarized along the sample plane, at an angle $\varphi$ with respect to X laboratory axis. By rotating the half-waveplate, we vary the orientation of the excitation linear polarization, and record the SHG emerging from the sample as function of the polarization angle φ. The x axis is taken parallel to the ZZ direction of the crystal and at angle $\theta$ from X. The y direction is then along the AC crystallographic direction, which coincides with the mirror symmetry axis (Fig. 1a).

The excitation field after passing the half-wave retardation plate can be expressed in laboratory coordinates by the Jones vector $\begin{pmatrix} E_0 cos\varphi \\ E_0 sin\varphi \end{pmatrix}$, where $E_0$ is the amplitude of the electric field. The expression of this vector in crystal coordinates is given by multiplying with the rotation matrix $\begin{pmatrix} cos\theta & sin\theta \\ -sin\theta & cos\theta \end{pmatrix}$, giving $E^\omega = \begin{pmatrix} E_0 cos(\varphi - \theta) \\ E_0 sin(\varphi - \theta) \end{pmatrix}$.

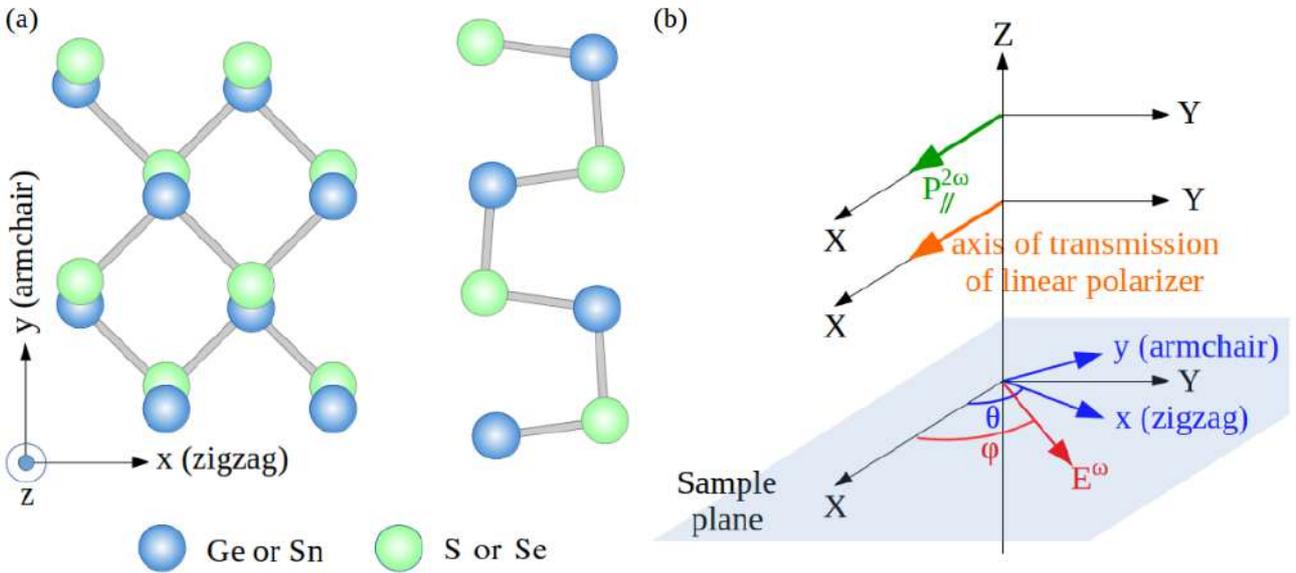

**Fig. 1** | (a) Schematic of the crystal structure of orthorhombic 2D MXs, as seen from the top (left) and the side (right). (b) Illustration of the two coordinate systems, the laboratory X, Y, Z, and the crystal, x, y, Z, adopted in our experimental configuration. The angles φ and θ describe the orientation of the laser field $E^\omega$ and the ZZ crystallographic direction relative to the X laboratory axis, respectively. $P_{//}^{2\omega}$ shows the detected component of the generated SHG field.

The 2D MXs belong to the non-centrosymmetric, orthorhombic point group C$_{2v}$ (mm2) [13]. Thus, they have five independent, nonzero SHG susceptibility tensor elements, namely: $\chi^{(2)}_{yxx}, \chi^{(2)}_{yyy}, \chi^{(2)}_{yzz}, \chi^{(2)}_{xyx} = \chi^{(2)}_{xxy}$, and $\chi^{(2)}_{zzy} = \chi^{(2)}_{zyz}$, where $\chi^{(2)}_{ijk}$ is the second-order nonlinear optical susceptibility tensor element along the different directions [13]. We note that these are the nonzero $\chi^{(2)}_{ijk}$ elements reported in [17], with the substitution $[x, y, z] \to [y, z, x]$. As a result, the nonlinear polarization can be written in matrix form as [17]:

$$\begin{pmatrix} P_x^{2\omega} \\ P_y^{2\omega} \\ P_z^{2\omega} \end{pmatrix} = \varepsilon_0 \begin{pmatrix} 0 & 0 & 0 & 0 & 0 & \chi^{(2)}_{xxy} \\ \chi^{(2)}_{yxx} & \chi^{(2)}_{yyy} & \chi^{(2)}_{yzz} & 0 & 0 & 0 \\ 0 & 0 & 0 & \chi^{(2)}_{zyz} & 0 & 0 \end{pmatrix} \begin{pmatrix} E_x^\omega E_x^\omega \\ E_y^\omega E_y^\omega \\ E_z^\omega E_z^\omega \\ 2E_y^\omega E_z^\omega \\ 2E_x^\omega E_z^\omega \\ 2E_x^\omega E_y^\omega \end{pmatrix} (1),$$

where $\varepsilon_0$ is the permittivity of the free space. Given that the excitation field is polarized along the sample plane, we have considered $E_z^\omega = 0$, and thus the SHG equation is reduced to:

$$\begin{pmatrix} P_x^{2\omega} \\ P_y^{2\omega} \end{pmatrix} = \varepsilon_0 E_0^2 \begin{pmatrix} \chi^{(2)}_{xxy} \sin[2(\varphi - \theta)] \\ \chi^{(2)}_{yxx} \cos^2(\varphi - \theta) + \chi^{(2)}_{yyy} \sin^2(\varphi - \theta) \end{pmatrix} (2),$$

where terms including only three independent SHG susceptibility tensor elements survive. We then transform this expression back to laboratory coordinates. In order to account for the effect of the linear polarizer placed before the detector, we multiple the SHG field with the Jones matrix $\begin{pmatrix} \cos^2\zeta & \sin\zeta\cos\zeta \\ \sin\zeta\cos\zeta & \sin^2\zeta \end{pmatrix}$, where ζ is the angle between the transmission axis of the polarizer and the X laboratory axis. In this work, we have set $\zeta = 0$, i.e., the axis of transmission of the polarizer parallel to X axis, and we measure the corresponding component of the SHG response, $P_{//}^{2\omega}$, whose intensity $I_{//}^{2\omega}$ is calculated as:

$$I_{//}^{2\omega} \sim \frac{1}{16} \left[ \begin{array}{c} 2(\chi^{(2)}_{yxx} + \chi^{(2)}_{yyy})\sin\theta + (2\chi^{(2)}_{xxy} - \chi^{(2)}_{yxx} + \chi^{(2)}_{yyy})\sin(\theta - 2\varphi) \\ + (2\chi^{(2)}_{xxy} + \chi^{(2)}_{yxx} - \chi^{(2)}_{yyy})\sin(3\theta - 2\varphi) \end{array} \right]^2 \quad (3)$$

In this relationship, the SHG intensity is expressed in terms of the absolute values of the χ$^{(2)}$ tensor elements. Instead, we can express it in terms of dimensionless ratios of the χ$^{(2)}$ tensor elements, obtaining:

$$I_{//}^{2\omega} = a[2(b+1)\sin\theta + (2c - b + 1)\sin(\theta - 2\varphi) + (2c + b - 1)\sin(3\theta - 2\varphi)]^2 (4),$$

where

$$b = \chi^{(2)}_{yxx}/\chi^{(2)}_{yyy}, c = \chi^{(2)}_{xxy}/\chi^{(2)}_{yyy} (5),$$

and $a = \varepsilon_0^2 E_0^4 / \left[16\left(\chi^{(2)}_{yyy}\right)^2\right]$ is a multiplication factor. The SHG intensity can also be expressed in the equivalent form, which is used to fit the P-SHG experimental data:

$$I_{//}^{2\omega} = a[2(b+1)\sin\theta + (b-1)[(e-1)\sin(\theta - 2\varphi) + (e+1)\sin(3\theta - 2\varphi)]]^2 (6),$$

where

$$e = 2c/(b - 1) (7).$$

In Fig. 2, we present the numerical simulation of the $I_{//}^{2\omega}$ modulation, described by Eq. 4, in polar plots, as a function of the orientation of the linear polarization of the excitation field, φ, for fixed values of b, c, and for different AC/ZZ directions. A video presenting the P-SHG polar diagrams with a step of 10° in the ZZ direction θ is shown in Supplementary information. Remarkably, the shape of the polar diagram itself is predicted to change for different values of the AC/ZZ directions. Three possible shapes are obtained: one with four symmetric lobes, one with four lobes symmetric in pairs, and one with two symmetric lobes. This shape change is in contrast to the corresponding behavior of the P-SHG polar diagrams of monolayer TMDs, which belong to the $D_{3h}$ point symmetry group, for which the $\chi^{(2)}$ tensor exhibits only one independent element [31, 32]. In that case, we have observed a characteristic four-lobe pattern, which rotates for different values of the crystal armchair direction [15, 16] (see also a video in the Supplementary information).

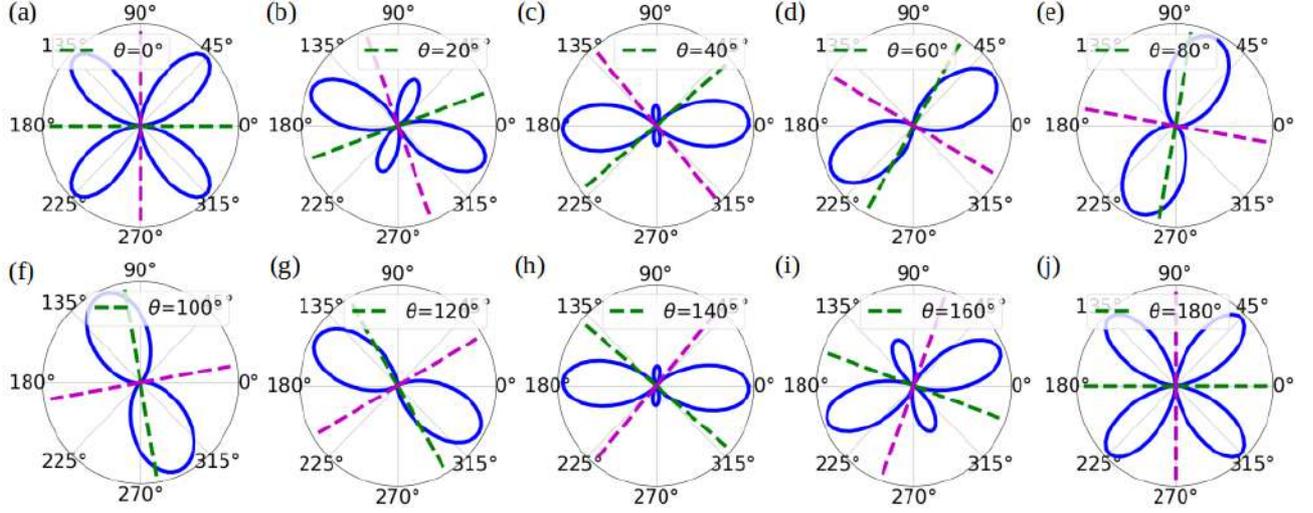

**Fig. 2** | Numerical simulations of the theoretical P-SHG intensity produced by a 2D MX, described by Eq. 4. We plot $I_{//}^{2\omega}$ in polar diagrams, as a function of the orientation of the linearly polarized excitation angle φ for fixed values $b = 5$, $c = 6.5$, and for different ZZ directions θ, with θ ϵ [0°, 180°] with step 20°. The AC/ZZ directions are illustrated with the magenta/green lines for each case.

The changes in the shape of the P-SHG polar diagrams, shown in Fig. 2, reflect the in-plane anisotropy of the orthorhombic MXs. Indeed, the origin of this shape change is described by Eq. 6, where the SHG intensity depends on four parameters, i.e., a, b, c and θ. We are therefore able to establish a direct link between the P-SHG intensity modulation and the in-plane anisotropy of orthorhombic MXs through these four parameters. In particular, the shape of the theoretically predicted P-SHG polar diagrams shown in Fig. 2 is determined by the corresponding ZZ direction θ and the tensor elements ratios b and c. The parameters b and c (Eq. 5) denote the relative contribution of different directions to the SHG signals. In Figs. S1, S2 in Supplementary information we simulate the effect of different values of the $\chi^{(2)}$ element ratios b and c on the P-SHG polar diagrams.

In contrast to the monolayer TMDs where the AC direction can be calculated modulo 60° (due to their threefold rotational symmetry i.e., the fact that they have three equivalent AC axes), in the case of 2D MXs, the AC direction is unique. This is readily reflected in the SHG polar diagrams of MXs which are the same every 180° in the AC/ZZ direction.

To describe the SHG intensity generated from a few-layer orthorhombic 2D MX, with N number of layers, we extend the interference model introduced for 2D TMDs [33, 38]. Neglecting propagation effects, the second harmonic field arising will have the form of vector superposition:
$$\boldsymbol{E}^{2\omega} = \boldsymbol{E}_1^{2\omega} + \boldsymbol{E}_2^{2\omega} + \ldots + \boldsymbol{E}_N^{2\omega} \quad (8),$$
where the indices denote the second harmonic signal from the corresponding layers. The total SHG intensity produced by the N-layer structure, will then be:
$$I^{2\omega} = |\boldsymbol{E}_1|^2 + |\boldsymbol{E}_2|^2 + \ldots + |\boldsymbol{E}_N|^2 + 2\boldsymbol{E}_1 \cdot \boldsymbol{E}_2 + \ldots + 2\boldsymbol{E}_{N-1} \cdot \boldsymbol{E}_N \quad (9)$$
$$I^{2\omega} = I_1 + I_2 + \ldots + I_N + 2\sqrt{I_1 I_2} \cos\delta_{1,2} + \ldots + 2\sqrt{I_{N-1} I_N} \cos\delta_{N-1,N} \quad (10) ,$$

where $\delta_{i,j}$, $i,j = 1,2,\ldots,N$ denote the relative angle between layers $i$ and $j$, i.e., the twist-angles, and the frequency index $2\omega$ is suppressed for simplicity. If we assume for simplicity that the SHG intensity from the individual layers is equal ($I_1 = I_2 = \ldots = I_N = I_{ML}$) and that the three layers are aligned (i.e., all twist-angles are zero), we find that

$$I^{2\omega} = NI_{ML} + I_{ML}N(N-1) \quad (11)$$
$$I^{2\omega} = N^2 I_{ML} \quad (12).$$

This is the well-known result that the SHG intensity from 2D flakes with zero twist-angle scales quadratically with the number of layers.

## 2.2 Nonlinear microscope

Our experimental setup is based on an inverted microscope (Axio Observer Z1, Carl Zeiss), which uses a fs laser (FLINT FL1 Yb Oscillator, ~6 W, 1028 nm, ~76 MHz, 30 fs, Light Conversion) to pump nonlinear optical processes (Fig. 3). A pair of silver-coated galvanometric (galvo) mirrors (6215H, Cambridge Technology) guides the laser beam into the microscope, allowing to raster-scan stationary samples. The beam passes through a zero-order half-wave retardation plate (QWPO-1030-10-2, CVI Laser), which is placed in a motorized rotation stage (M-060.DG, Physik Instrumente), with which we can rotate the linear polarization of the excitation field with accuracy of 0.1°. A pair of achromatic lenses suitably expands the beam diameter to fill the back aperture of the objective lens (Plan-Apochromat × 40/1.3 NA, Carl Zeiss).

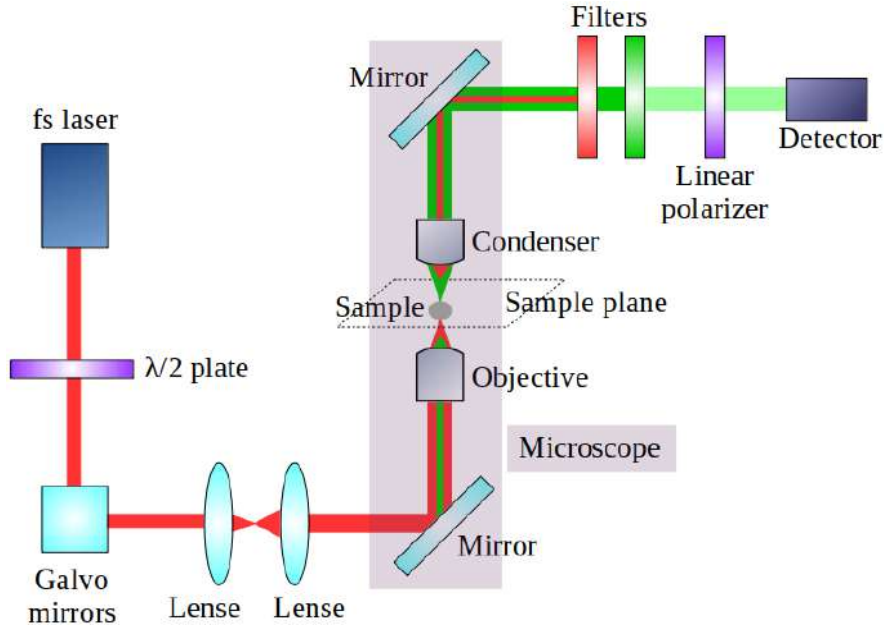

**Fig. 3** | Illustration of the experimental setup. The fs laser beam is guided into the microscope and excites SHG produced by a stationary 2D SnS crystal. By rotating a λ/2 plate, we rotate the orientation of the linear polarization of the excitation field, as a function of which we record the second harmonic signal. A pair of galvanometric mirrors is used to raster-scan an area of the sample and obtain SHG images.

At the motorized turret box of the microscope, we have the option of using either a silver mirror or a dichroic mirror, both at 45° just below the objective, depending on whether we wish to collect the produced signal in the forward or backwards (epi) detection geometry, respectively. In this work, we collect the signal in the forward direction, using the silver mirror which is insensitive to the laser beam polarization. The objective lens tightly focuses the beam onto the stationary sample that produces SHG, which is collected by a condenser lens (achromatic-aplanatic, 1.4 NA, Carl Zeiss). We then use suitable short-pass (FF01-680/SP, Semrock) and narrow bandpass (FF01-514/3,

Semrock) filters to cut off residual laser light and any other unwanted signal. Finally, a linear polarizer (LPVIS100-MP, ThorLabs) is placed before the detector which is based on a photomultiplier tube module (H9305-04, Hamamatsu), in order to select the detected SHG polarization.

The galvanometric mirrors and the photomultiplier tubes are connected to a connector block (BNC-2110, National Instruments Austin), which is interfaced to a PC through a DAQ (PCI 6259, National Instruments). The coordination of the detector recordings with the galvanometric mirrors for the image formation, as well as the movement of the motors, is carried out using LabView (National Instruments).

This setup allows us to record spatially resolved SHG intensity images from a sample region, while rotating the linear polarization of the excitation field, performing P-SHG imaging. Each image (of 500x500 pixels in this work), corresponds to a sample area of size from a few μm to hundreds of μm, depending on how we have set the movement of the galvo mirrors. We also note that our diffraction-limited spatial resolution is approximately 400 nm (NA=1.3, $\lambda_{exc}$=1028nm). For the data analysis, we used the MATLAB (The Mathworks, Inc) programming language [39], the open-source Python programming language [40] and the open-source ImageJ image analysis software [41].

## 2.3 Sample preparation and characterization

The ultrathin layers of SnS sheets are isolated from bulk crystal via liquid phase exfoliation (LPE) [42-44] (see Supplementary information). Fig. S4a represents the UV-vis extinction spectra of isolated SnS sheets in acetone. The extinction spectra of SnS exhibited a broad absorption profile with UV-vis-NIR spectral range, accompanied with a shoulder at 420 nm. Such extinction spectral feature is consistent with the reported thin layers of SnS sheets [42-44], which suggests the isolation of an ultrathin layer of SnS sheets. In addition, atomic force microscopy (AFM) images of isolated SnS are collected to acquire the flake dimensionality. As may be seen from a representative area of our sample, shown in Fig. S4b, the height profile of each SnS sheet is varying, with all being less than ~1.1 nm. Given that the monolayer SnS thickness is ~0.6 nm and that AFM measurements of liquid phase exfoliated SnS have been reported to overestimate the thickness because of the solvent overlayer [43], our sample is identified to contain monolayers and bilayers.

## 3. Results and discussion

Using our custom-built polarization-resolved nonlinear microscope presented in Fig. 3, we raster-scan a specific sample area, and by rotating the linear polarization angle of the pump beam, φ, with a step of 2°, we record 180 spatially resolved images of $I_{//}^{2\omega}$. We then fit those images with Eq. 6 and estimate: i) the ZZ direction θ (and thus the perpendicular AC direction too); ii) the $\chi^{(2)}$ tensor element parameters b and e; and iii) the multiplication factor a. This is performed for every pixel of the image enabling the extraction of spatially resolved images of the ZZ crystal direction and the tensor element parameter values, as well as the corresponding distributions of such values.

In order to confirm the SHG process, we measure the average SHG intensity produced by a 2D SnS flake as function of the excitation power, shown in Fig. S5 in a log-scale plot. Indeed, we obtain a quadratic power-law dependence, as expected. In Fig. 4, we present representative P-SHG images of different 2D SnS crystals, belonging in the same field of view, for several orientations of the laser linear polarization φ denoted by the red arrow. A video of all the 180 P-SHG images obtained, with φ ϵ [0°, 360°] with step 2°, can be found in the Supplementary information. A number of SnS flakes, appearing as bright spots of submicron dimensions, is observed. Six regions of interest (ROIs) containing SnS crystals are illustrated by the white arrows. While changing φ, we observe differences in the SHG intensity of the individual SnS flakes, in accordance with the theoretical predictions of Eqs. 4 and 6. We also note that the detected SHG signals from the different flakes are modulating out of phase. This is due to the different AC/ZZ crystallographic orientations and/or due to differences in the tensor element ratios among the different SnS flakes.

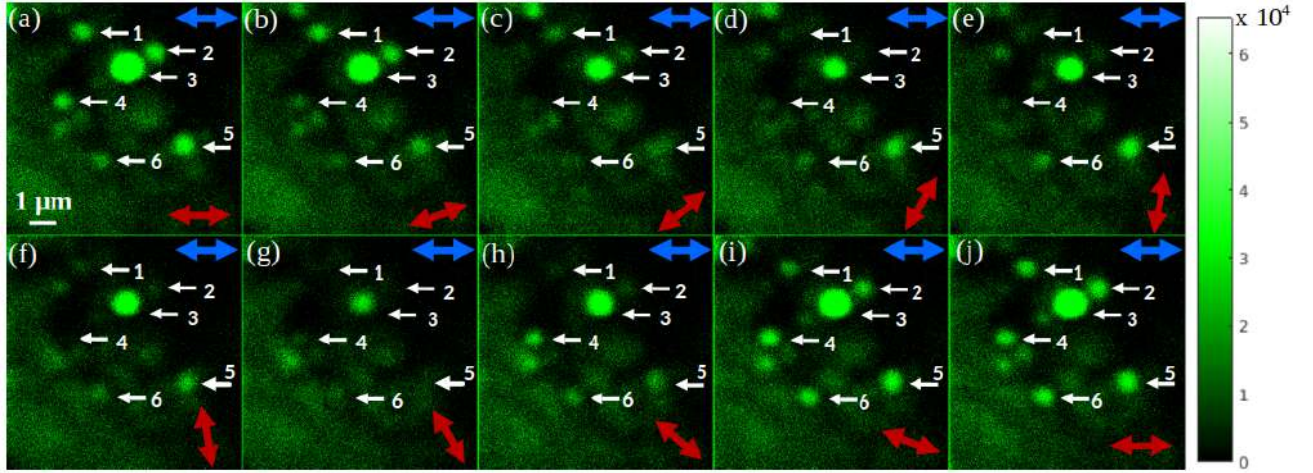

**Fig. 4** | Experimental P-SHG images of ultrathin SnS crystals belonging in the same field of view for different values of the orientation of the laser linear polarization (φ in Eq. 4), denoted by the red arrows, with φ ϵ [0°, 180°] with step 20°. The blue arrows indicate the direction of the polarization of the detected SHG signals. Brighter color indicates higher P-SHG intensity in arbitrary units. We note that the SHG signals from the 2D SnS crystals (ROIs 1-6 pointed by the arrows), are modulating out of phase. The scale bar in the first image illustrates 1 μm.

We focus on six regions of interest (ROIs) containing SnS crystals, illustrated by white arrows in Fig. 4. In Fig. 5a, we present for such ROIs the sum of the 180 P-SHG intensity images for all orientations of the excitation linear polarization φ. In Figs. 5b-g, we present polar plots of the P-SHG modulation (in red dots) taken from one pixel inside the ROIs depicted in Fig. 5a. We note that these diagrams confirm the theoretical prediction that different SnS flakes can produce P-SHG polar plots of different shape (see Fig. 2), depending on the ZZ crystallographic direction and the parameters b and e. Using our methodology, this is observed within the same field of view, providing new means of contrast. The third possible shape of the P-SHG polar diagrams, with two single lobes as theoretically predicted in Fig. 2d-g, has been also experimentally demonstrated for another SnS flake and is presented in Fig. S3. By fitting (blue line) the experimental data with Eq. 6, we are able to calculate the AC/ZZ crystallographic direction and the $\chi^{(2)}$ element ratios, i.e., parameters b and c for each pixel (summarized in Table 1).

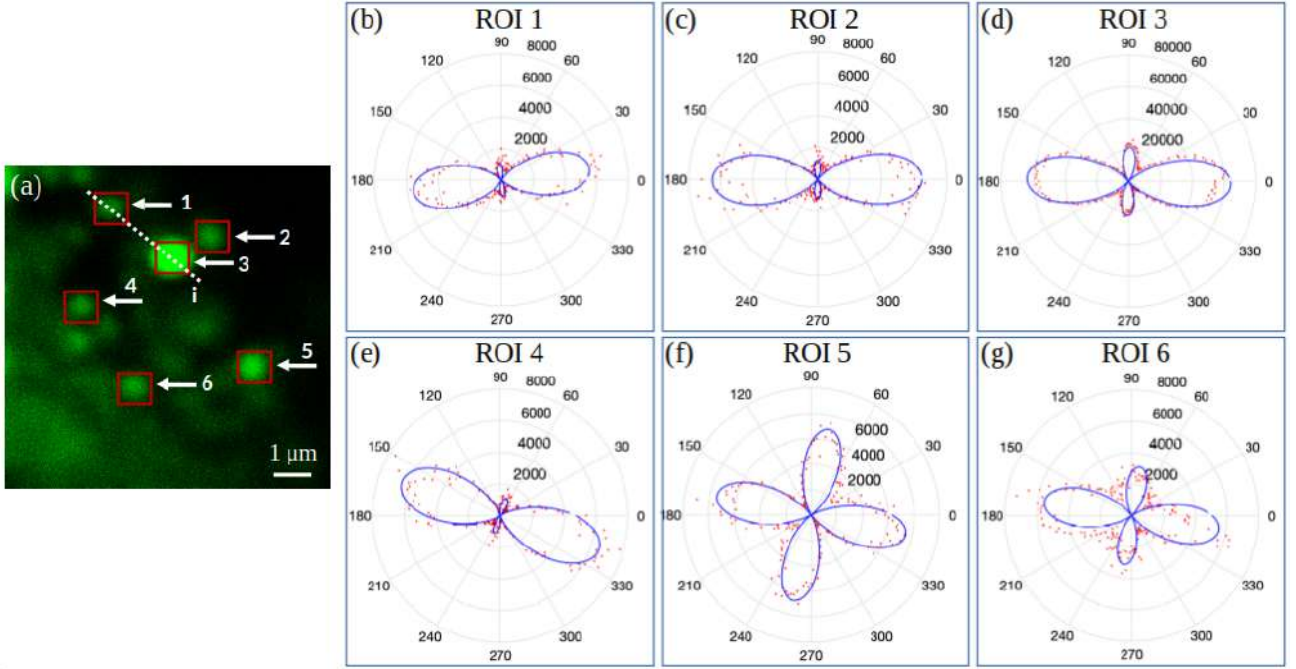

**Fig. 5** | (a) Sum of the SHG intensity for all orientations of the linearly polarized excitation angle φ, corresponding to the same field of view shown in Fig. 4. Brighter color indicates higher SHG intensity. The scale bar illustrates 1 μm. (b-g) Experimental data (in red dots) of the P-SHG intensity taken from one pixel inside each ROI depicted in Fig. 5a, presented in polar plots as function of the angle φ. By fitting (blue line) with Eq. 6, we are able to calculate the ZZ crystallographic direction and the tensor element parameters b and e, for each pixel (summarized in Table 1. Interestingly, the shape of the polar-diagrams changes for different flakes, which is the signature of differences in their in-plane anisotropy.

Table 1: Summary of the fitted parameters θ, a, b and e, for all datasets demonstrated in Fig. 5b-g, with the corresponding quality of fitting $R^2$. The $\chi^{(2)}$ parameter c is calculated through Eq. 7.

| ROI | Zigzag direction θ (°) | Parameter a (arb. units) | Parameter $b = \chi^{(2)}_{yxx}/\chi^{(2)}_{yyy}$ | Parameter e | Quality of fitting | Parameter $c = \chi^{(2)}_{xxy}/\chi^{(2)}_{yyy}$ |
|---|---|---|---|---|---|---|
| 1 | -33.96 | 2.78 | 5.05 | 7.33 | 85 % | 14.84 |
| 2 | -40.35 | 2.57 | 6.16 | 7.15 | 92 % | 18.45 |
| 3 | 40.76 | 2.29 | 16.34 | 8.55 | 91 % | 65.58 |
| 4 | 23.67 | 2.56 | 9.09 | 3.75 | 91 % | 15.17 |
| 5 | 29.25 | 0.30 | 13.05 | 3.77 | 73 % | 22.71 |
| 6 | 29.36 | 0.91 | 12.59 | 3.07 | 56 % | 17.79 |

This fitting can also be performed in a pixel-by-pixel manner, producing spatially resolved ZZ orientation maps. Such maps along with their corresponding image histograms are presented in Fig. 6, for the ROIs 1, 4-6 depicted in Fig. 5a. Although, to date, only micrometer-size SnS monolayers have been realized, our technique can provide useful information on crystal quality and the presence of grain boundaries and defects in larger-area crystals [31, 32]. Using the same fitting procedure, we additionally produce distributions of values (image histograms) for the $\chi^{(2)}$ tensor element parameters b and e, which are presented in Fig. 7. All such histograms are subsequently fitted with a Gaussian function in order to calculate the mean and standard deviation of the distribution of values for each parameter. Table 2 summarizes the results of the fitted parameters θ, b and e (mean and sigma), along with the values of c, calculated through Eq. 7. Considering the broad range of the b, c and e values, the results listed in Table 2 provide experimental and quantitative evidence on the highly anisotropic nature of the $\chi^{(2)}$ tensor of SnS.

To compare our findings with the literature, we use the values of the $\chi^{(2)}$ tensor elements of monolayer MXs that have been theoretically calculated from first-principles using density functional theory [13]. For the particular case of SnS and for excitation pulse centred at $1028 nm$ ($\hbar\omega \simeq 1.2 eV$), they have been calculated to be $\chi^{(2)}_{yyy} \simeq 65 \cdot 10^4 \frac{pm^2}{V}$, $\chi^{(2)}_{yxx} \simeq 50 \cdot 10^4 \frac{pm^2}{V}$, $\chi^{(2)}_{yyy} \simeq 10 \cdot 10^4 \frac{pm^2}{V}$ (see Fig. S2d in [13]). These values correspond to $b = \chi^{(2)}_{yxx}/\chi^{(2)}_{yyy} \simeq 5$, $c = \chi^{(2)}_{xxy}/\chi^{(2)}_{yyy} \simeq 6.5$, $e = 2\ c/(b-1) \simeq 3.25$, which agree with the experimental evidence that c is higher than b, while such values are within the same order of magnitude with the experimental values of Table 2. The deviations from the literature values, the broad histograms, and the different values of the $\chi^{(2)}$ parameters among different 2D SnS crystals may be attributed to i) deformation in the crystal lattice during the sample preparation [16] ii) varying contributions to the SHG signal from the $\chi^{(2)}$ tensor elements along different directions, and iii) the adopted fitting procedure.

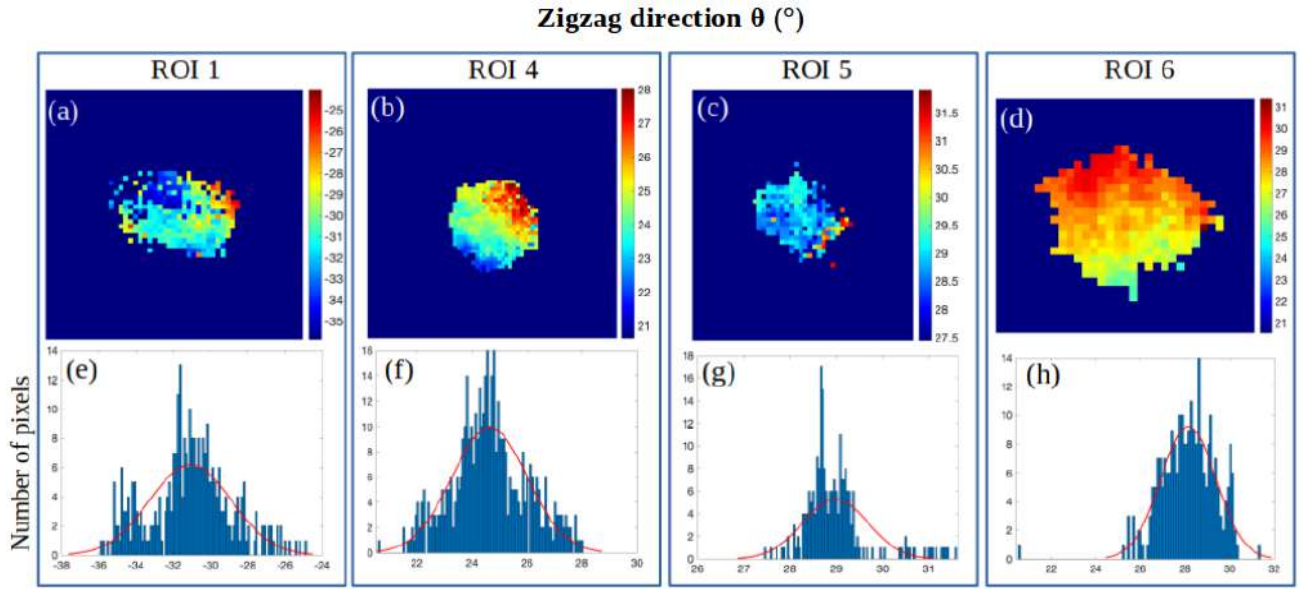

**Fig. 6** | (a-d) Pixel-by-pixel spatially resolved mapping of the ZZ crystallographic direction θ for the ultrathin SnS crystals which correspond to the ROIs 1, 4-6 depicted in Fig. 5a. We present pixels that survived quality of fitting larger than 80%, 88%, 67% and 50%, respectively. (e-h) Corresponding image histograms showing the distributions of the values of the ZZ directions and the Gaussian fit (red line). The fitted parameters of the Gaussian fit are summarized in Table 2.

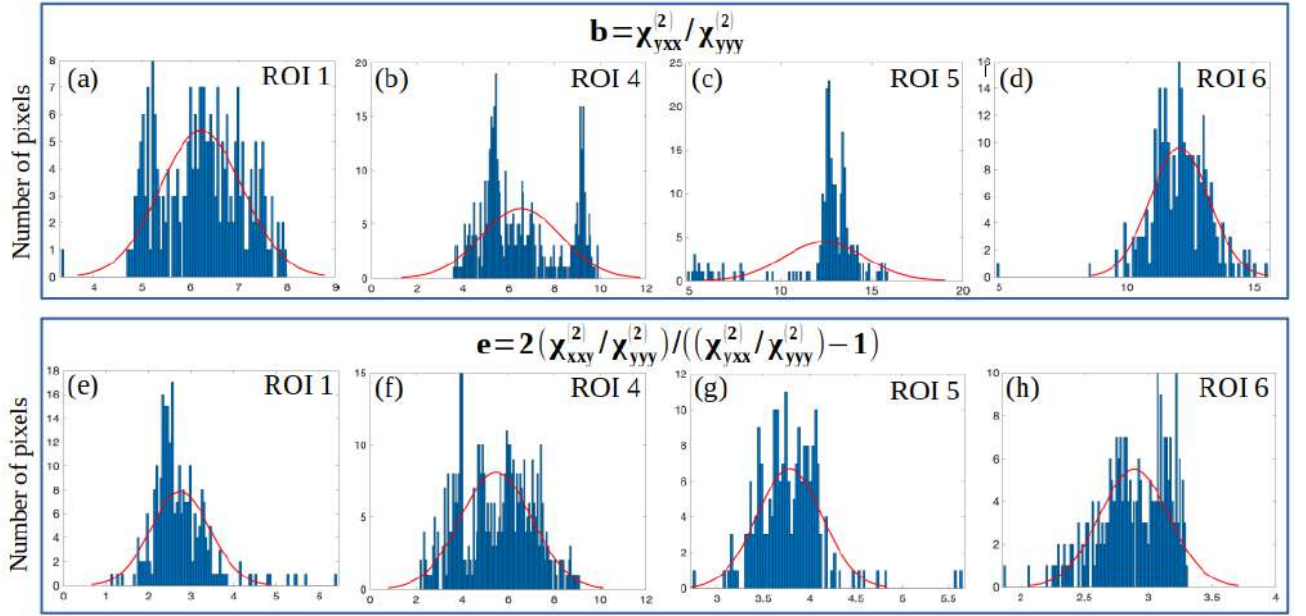

**Fig. 7** | (a-d) Histograms of the fitted $\chi^{(2)}$ parameter b (Eq. 5), and (e-h) the fitted $\chi^{(2)}$ parameter e (Eq. 7), for the ultrathin SnS crystals which correspond to the ROIs 1, 4-6 depicted in Fig. 5a. We present pixels that survived quality of fitting larger than 80%, 88%, 67% and 50%, respectively. The fitted parameters of the Gaussian fit (red line) are summarized in Table 2.

Table 2: Summary of the fitted parameters θ, b and e (mean and standard deviation), based on the Gaussian fit illustrated in the representative histograms in Fig. 6, 7. The results for all ultrathin SnS crystals which correspond to the ROIs shown in Fig. 5a are presented. The $\chi^{(2)}$ parameter c is calculated through the mean values of b and e using Eq. 7.

| ROI | Zigzag direction θ (°) | | Parameter $b = \chi^{(2)}_{yxx}/\chi^{(2)}_{yyy}$ | | Parameter $e = 2c/(b-1)$ | | Parameter $c = \chi^{(2)}_{xxy}/\chi^{(2)}_{yyy}$ | |
|---|---|---|---|---|---|---|---|---|
| | mean | standard deviation | mean | standard deviation | mean | standard deviation | mean | standard deviation |
| 1 | -31.07 | 2.19 | 6.22 | 0.86 | 2.74 | 0.69 | 7.15 | 1.08 |
| 2 | 0.41 | 35.50 | 12.33 | 5.81 | 3.23 | 2.48 | 18.30 | 8.45 |
| 3 | 38.22 | 1.55 | 19.73 | 2.75 | 5.17 | 1.32 | 48.42 | 7.13 |
| 4 | 24.63 | 1.36 | 6.53 | 1.74 | 5.46 | 1.58 | 15.10 | 3.23 |
| 5 | 28.98 | 0.70 | 12.39 | 2.20 | 3.78 | 0.35 | 21.53 | 2.31 |
| 6 | 28.15 | 1.24 | 12.08 | 1.15 | 2.89 | 0.27 | 16.01 | 1.12 |

We have also investigated the reason behind the considerably higher SHG intensity exhibited by one SnS flake (ROI 3) compared to its neighboring ones, as illustrated in Fig. 4, 5a. We note that, the integrated SHG intensity from each SnS flake shown in Fig. 5a (sum of all SHG intensities acquired for all the different excitation polarizations $\varphi \in [0°, 360°]$ with step 2°), is not polarization dependent anymore, and thus differences in the SHG intensities between the different flakes in the same image could solely attributed to differences in their number of layers. In order to quantify these differences, Table 3 summarizes the maximum SHG intensities detected in each ROI. We characterized as monolayer the flake with the minimum intensity, i.e., ROI 1. The number of layers $N$ for the other flakes is then determined using Eq. 12 ($N = \sqrt{I/I_{ML}}$). We note that ROIs 2, 4-6 could also be characterized as monolayers, while ROI 3 as a trilayer with zero twist-angle. Fig. 8 shows the SHG intensity along the dashed line i shown in Fig. 5a, confirming that ROI 3 exhibits SHG intensity approximately nine times higher.

Table 3: Summary of the maximum SHG intensity detected for the ultrathin SnS crystals which correspond to the ROIs shown in Fig. 5a. As monolayer is characterized the flake with the minimum intensity, i.e., ROI 1, while the number of layers $N$ for the other flakes is determined using Eq. 12.

| ROI | Max. SHG intensity (x $10^5$) (arb. units) | Number of layers N |
|---|---|---|
| 1 | 4.7 | 1 |
| 2 | 5.4 | 1.1 |
| 3 | 49 | 3.2 |
| 4 | 5.6 | 1.1 |
| 5 | 8.3 | 1.3 |
| 6 | 5.9 | 1.1 |

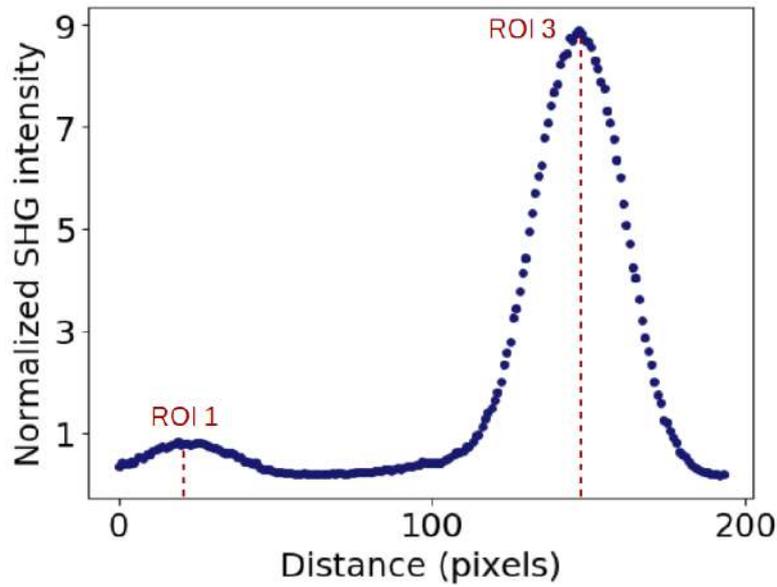

**Fig. 8** | (a) SHG intensity along the dashed line i shown in Fig. 5a, normalized to the maximum intensity in ROI 1. The intensity in ROI 3 is found to be nine times higher than that of ROI 1, implying that ROI 3 could be a SnS trilayer.

**4. Conclusions**

In summary, taking advantage of the orthorhombic crystal structure of 2D SnS crystals that induces SHG conversion, we have optically mapped the in-plane optical anisotropy of 2D SnS flakes with high resolution. By performing P-SHG imaging microscopy, we found that the P-SHG polar plot changes shapes among different flakes. This finding reflects the effect of the in-plane anisotropy of the orthorhombic MXs on their nonlinear optical properties. This is demonstrated for different 2D SnS flakes belonging to the same field of view. Our approach provides new means of contrast that discriminates 2D SnS flakes in the same image based on their in-plane anisotropy. By fitting the experimental data with a nonlinear optics model, that accounts for the material crystal structure, we were able to calculate and map with high resolution the AC/ZZ crystallographic orientation of each flake, and to estimate two second-order nonlinear optical susceptibility tensor element ratios for every sample point. This methodology can be used to spatially determine in large crystal areas the optical in-plane anisotropy in different orthorhombic MXs crystals. Our results provide a novel, all-optical probe of the in-plane anisotropic properties of orthorhombic MXs based on nonlinear optics, that is useful for emerging fundamental studies and optoelectronic applications of these materials.

**Author contributions**

ES and GK guided the research. SP and GMM conducted the optical experiments. GMM, LM and SP developed the theoretical model. SP, GMM, LM and AL conducted the data analysis. ASS prepared and characterized the samples. AL provided technical support. All authors contributed to the discussion and preparation of the manuscript.

**Data availability**

The data that support this study are available from the corresponding authors upon reasonable request.

**Acknowledgments**

This research is co-financed by Greece and the European Union (European Social Fund-ESF) through the Operational Programme "Human Resources Development, Education and Lifelong Learning 2014-2020" in the context of the project "Crystal quality control of two-dimensional materials and their heterostructures via imaging of their non-linear optical properties" (MIS 5050340).

**Conflict of interest**
The authors declare that they have no conflict of interest.

# Supplementary information

# Nonlinear optical imaging of in-plane anisotropy in two-dimensional SnS

G. M. Maragkakis[1,2+], S. Psilodimitrakopoulos[1+], L. Mouchliadis[1] A. S. Sarkar[1], A. Lemonis[1], G. Kioseoglou[1,3], and E. Stratakis[1,2*]

[1]Institute of Electronic Structure and Laser, Foundation for Research and Technology-Hellas, Heraklion Crete 71110, Greece.

[2]Department of Physics, University of Crete, Heraklion Crete 71003, Greece.

[3]Department of Materials Science and Technology, University of Crete, Heraklion Crete 71003, Greece.

* Correspondence: E. Stratakis, e-mail: **stratak@iesl.forth.gr**

[+] These authors contributed equally to this work

**Simulation of the P-SHG intensity**

In Fig. S1, S2 we present the effect of different values of the second-order nonlinear optical susceptibility tensor element ratios b and c, respectively, on the polarization-resolved second harmonic generation (P-SHG) intensity produced by a group IV monochalcogenide, described by Eq. 4.

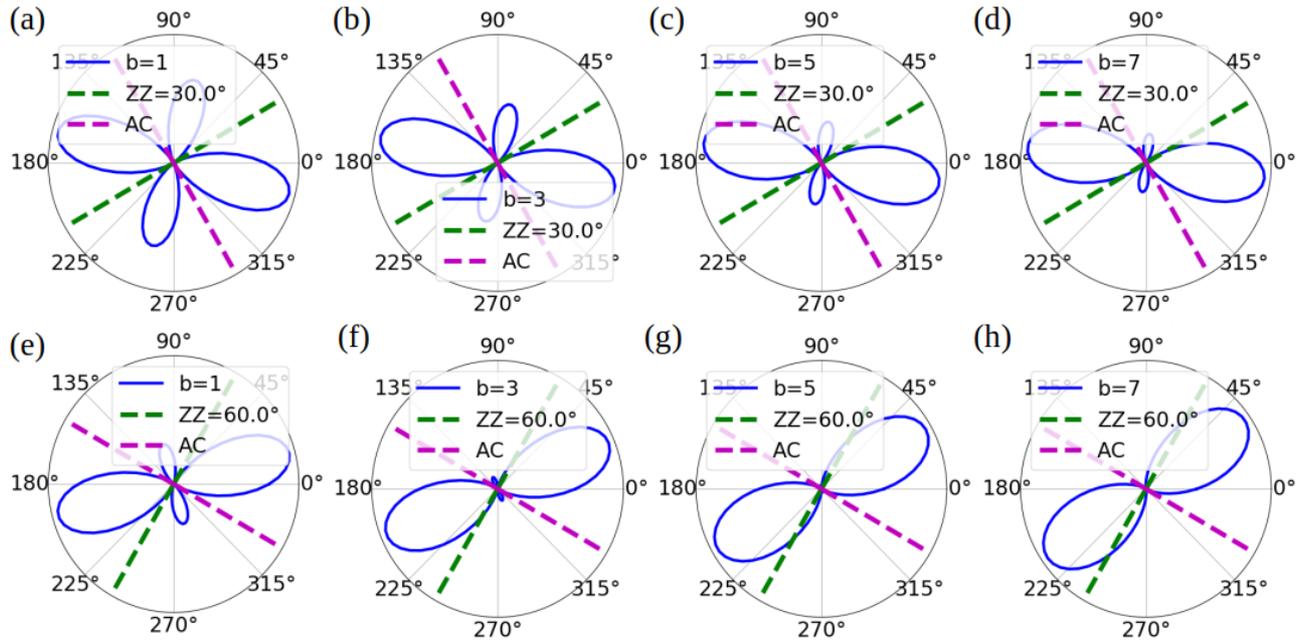

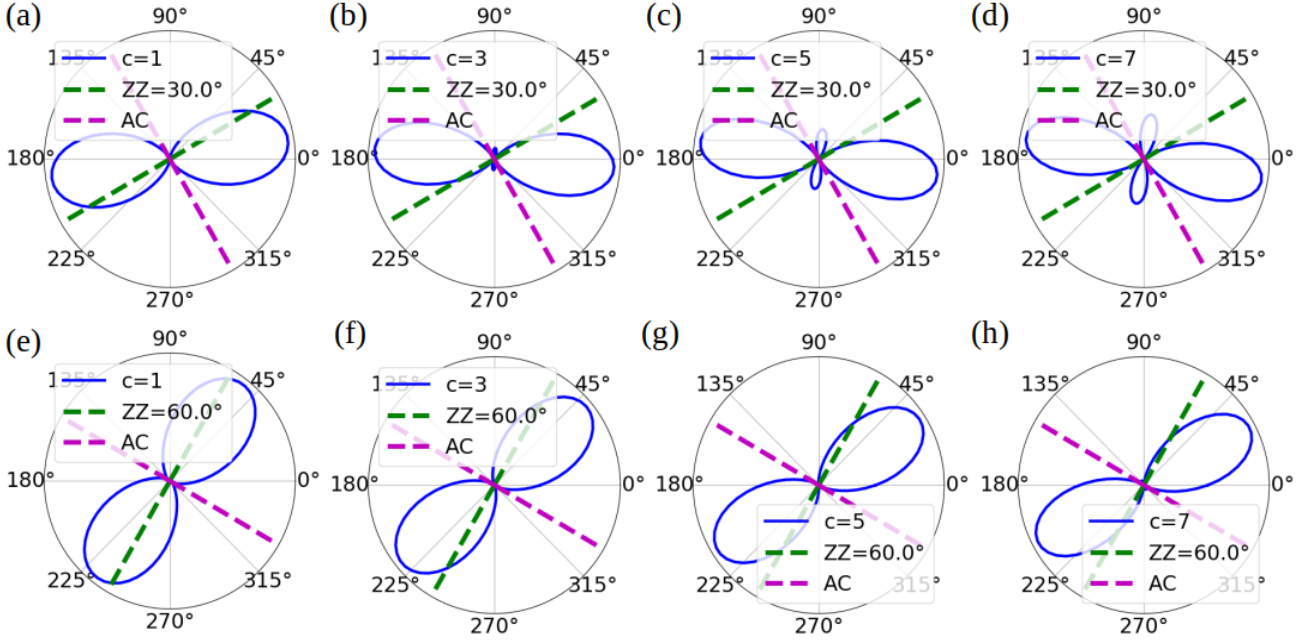

**Fig. S1** | Theoretical simulation of the P-SHG intensity produced by a group IV monochalcogenide, described by Eq. 4. In order to investigate the effect of the x$^{(2)}$ tensor element ratio b, we plot $I_{//}^{2\omega}$ in polar diagrams, as function of the orientation of the linear polarization of the pump laser beam, φ, for fixed value $c = 6.5$, and for two different armchair(AC)/zigzag(ZZ) directions.

**Fig. S2** | Theoretical simulation of the P-SHG intensity produced by a group IV monochalcogenide, described by Eq. 4. In order to investigate the effect of the x$^{(2)}$ tensor element ratio c, we plot $I_{//}^{2\omega}$ in polar diagrams, as function of the orientation of the linear polarization of the pump laser beam, φ, for fixed value $b = 5$, and for two different AC/ZZ directions.

**P-SHG polar diagram with two lobes**
In Fig. S3 we present a P-SHG polar diagram from a flake of similar SnS, which exhibits the third possible shape predicted by Eq. 4, i.e. with two lobes.

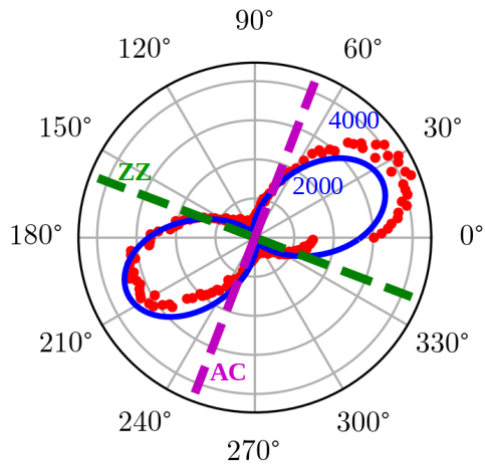

**Fig. S3** | Experimental data (in red) of the SHG intensity as function of the incident field polarization angle, φ, in polar plots, for a flake of similar SnS. By fitting (blue line) with the SHG equation we have derived (Eq. 4), we are able to calculate the AC/ZZ crystallographic direction for each flake.

**Preparation of SnS sheets**
In order to prepare the SnS sheets, we follow our previous work on liquid phase exfoliation of SnS. In brief, tin (II) sulfide granular trace metals basis (Sigma Aldrich) was dissolved in acetone (3 mg/ml) under nitrogen ($N_2$) atmosphere. The solution was ultrasonicated for 20 hours under 80 W power and 37 kHz frequency (Elma Schmidbauer GmbH). The sonicated dispersion was centrifuged at 8000 rpm for 15 min. The suspended SnS layer was isolated from the bulk precipitate. The collected SnS suspension was utilized for further experiments.

**Extinction spectroscopy**
Extinction spectra of the isolated SnS dispersions were acquired with a Perkin Elmer, Lamda 950 UV/VIS/NIR spectrometer.

**Atomic force microscopy**
Tapping mode atomic force microscopy (Multimode AFM from Digital Instruments, Bruker) was employed to capture the surface topography image of isolated SnS flakes. The sample was prepared by drop casting the SnS dispersion on cleaned Si substrate. The sample was then dried and placed on the microscope stage to scan.

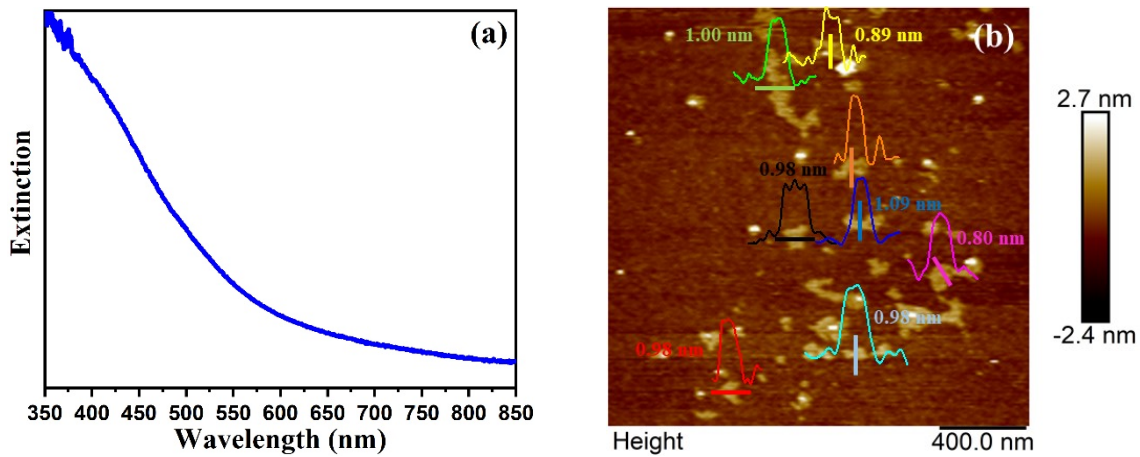

**Fig. S4** | (a) UV-Vis extinction spectra of liquid phase exfoliated SnS sheets. (b) Atomic force microscopic topography image of SnS sheets with flake height. Scale bar is 400 nm.

**Power-law dependence of SHG**
In Fig. S5 we present the power-law dependence of SHG, by measuring the average SHG intensity produced by a 2D SnS flake as function of the excitation power. The quadratic dependence confirms the SHG process.

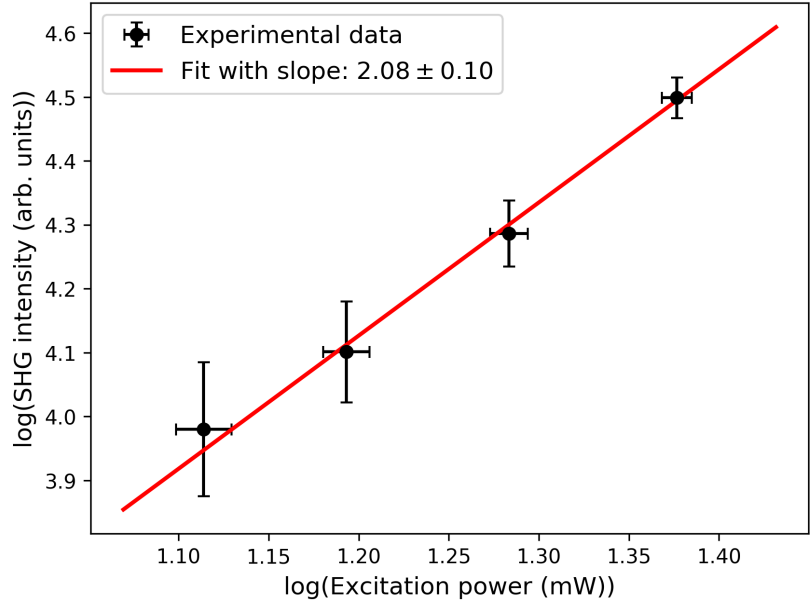

Fig. S5 | Log-scale plot of the average SHG intensity produced by a 2D SnS flake as function of the excitation power. The quadratic dependence confirms the SHG process

**Description of Video 1**

We present a video of theoretical simulations of the P-SHG intensity produced by a group IV monochalcogenide monolayer, described by Eq. 4. The second harmonic signal is plotted in polar diagrams as function of the orientation of the laser linear polarization, φ, for fixed values $b = 5$, $c = 6.5$, and for the zigzag (ZZ) direction, θ, changing with step $10^0$.

**Description of Video 2**

We present a video of theoretical simulations of the P-SHG intensity produced by a transition metal dichalcogenide monolayer. The second harmonic signal is plotted in polar diagrams as function of the orientation of the laser linear polarization, and for the armchair direction (AC) changing with step $5^0$.

**Description of Video 3**

We present a video of the P-SHG imaging of ultrathin SnS. We record the second harmonic generation signal while rotating the orientation of the laser linear polarization with step $2^0$. We have chosen a contrast for which the brighter area is saturated and the rest are more visible.

**Author contributions**
ES and GK guided the research. SP and GMM conducted the optical experiments. GMM, LM and SP developed the theoretical model. SP, GMM, LM and AL conducted the data analysis. ASS prepared and characterized the samples. AL provided technical support. All authors contributed to the discussion and preparation of the manuscript.

**Data availability**
The data that support this study are available from the corresponding authors upon reasonable request.

**Acknowledgments**
This research is co-financed by Greece and the European Union (European Social Fund-ESF) through the Operational Programme "Human Resources Development, Education and Lifelong Learning 2014-2020" in the context of the project "Crystal quality control of two-dimensional materials and their heterostructures via imaging of their non-linear optical properties" (MIS 5050340).